# Correlation between Fermi surface reconstruction and superconductivity in pressurized FeTe$_{0.55}$Se$_{0.45}$

\#

Gongchang Lin[1,3]*, Jing Guo[1,4]*, Yanglin Zhu[2], Shu Cai[1], Yazhou Zhou[1], Cheng Huang[1,3], Chongli Yang[1], Sijin Long[1,3], Qi Wu[1], Zhiqiang Mao[2], Tao Xiang[1,3] and Liling Sun[1,3,4]†

[1]*Institute of Physics, National Laboratory for Condensed Matter Physics, Chinese Academy of Sciences, Beijing, 100190, China*

[2]*Department of Physics, Pennsylvania State University, University Park, Pennsylvania 16802, USA*

[3]*University of Chinese Academy of Sciences, Department of Physics, Beijing 100190, China*

[4]*Songshan Lake Materials Laboratory, Dongguan, Guangdong 523808, China*

\#

Here we report the first results of the high-pressure Hall coefficient ($R_H$) measurements, combined with the high-pressure resistance measurements, at different temperatures#on the putative topological superconductor FeTe$_{0.55}$Se$_{0.45}$. We find the intimate correlation of sign change of $R_H$, a fingerprint to manifest the reconstruction of Fermi surface, with structural phase transition and superconductivity. Below the critical pressure ($P_c$) of 2.7 GPa, our data reveal that the hole - electron carriers are thermally balanced ($R_H$=0) at a critical temperature ($T^*$), where $R_H$ changes its sign from positive to negative, and concurrently a tetragonal-orthorhombic phase transition takes place. Within the pressure range from ambient pressure to $P_c$, $T^*$ is continuously suppressed by pressure, while $T_c$ increases monotonically. At about $P_c$, $T^*$ is undetectable and $T_c$ reaches a maximum value. Moreover, a pressure-induced sign change of $R_H$ is found at ~ $P_c$ where the orthorhombic-monoclinic phase transition occurs. With further compression, $T_c$ decreases and disappears at ~ 12 GPa. The correlation among the electron-hole balance, crystal structure and superconductivity found in the pressurized FeTe$_{0.55}$Se$_{0.45}$ implies that its nontrivial superconductivity is closely associated with its exotic normal state resulted from the interplay between the reconstruction of the Fermi surface and the change of the structural lattice.

The discovery of Fe-based superconductors provides a new platform not only for understanding the microscopic mechanism of high-temperature superconductivity beyond the copper oxide superconductors [1,2], but also for finding new phenomena from correlated electron systems. Among Fe-based superconductors, iron selenide (FeSe) is distinct; it has the simplest crystal structure [3] and shows sensitive effect of pressure on the superconducting transition temperature ($T_c$) [4,5]. The isovalent substitution Se with Te in FeSe superconductors can increase $T_c$ from 8 K to about 15 K [6-8], and more attractively, an unusual interplay between the resonance and the incommensurate magnetism has been found only in the crystals with an average composition near FeTe$_{0.5}$Se$_{0.5}$ [7,9,10]. Intriguingly, recent high-resolution angle resolved photoelectron spectroscopy and scanning tunneling spectroscopy experiments find the evidences for Dirac-cone type spin-helical surface states [11] and Majorana bound states in FeTe$_{0.55}$Se$_{0.45}$ superconductor [12], which is a signature of topological superconductivity. These new findings have renewed research interests of this material. One particularly interesting direction is to explore the variation of its electronic state with lattice structure. Results from such work are expected to reveal insights into the nature of the topological superconductivity of this material.

In general, the unconventional superconductivity of a given material is dictated by multiple degrees of freedom of charge, spin, orbital and lattice. These degrees of freedom as well as the interactions among them can be manipulated by control parameters such as pressure, magnetic field and chemical doping [13-18]. Pressure tuning is a clean way to provide significant information on co-evolution among superconductivity, electronic state and crystal structure without changing the chemistry, and to result in a deeper understanding on the underlying physics of the exotic state emerging from ambient-pressure materials. In this study, we performed *in-situ* high pressure transport measurements on the high quality single crystals of FeTe$_{0.55}$Se$_{0.45}$, with the attempt to find such kind of co-evolution information.

The single crystals with nominal composition of FeTe$_{0.55}$Se$_{0.45}$ were grown using a flux method [19]. The values of the midpoint $T_c$s of the samples from the two batches were determined to be 13.5 K and 13.7.8 K, respectively (Fig.S1). High pressure was generated by a diamond anvil cell made of BeCu alloy with two opposing anvils. A four-probe method was applied for our resistance measurements. Diamond anvils with 300 μm and 400 μm culets (flat area of the diamond anvil) were used for several independent measurements. In the experiments, we employed platinum foil as electrodes, rhenium plate as gasket, cubic boron nitride as insulating material and NaCl as pressure medium. High-pressure Hall coefficient was measured through Van der Pauw method under magnetic field generated from a superconducting coil [see Fig.S2 in the SI]. In the measurements, the contacts for current (I) and voltage (V) are swapped for positive and negative fields. Pressure in all measurements is determined by the ruby fluorescence method [20].

Figure 1 displays the temperature dependence of electrical resistance at different pressures. We find that the superconducting transition temperature ($T_c$) of sample 1 increases upon elevating pressure and then decreases upon further compression (Fig.1a and 1b), in good agreement with the results reported previously [21-26]. Similar results were obtained in the measurements on sample 2 (Fig.1c and 1d), *i.e.* $T_c$ first shows an increase in the low pressure range, reach a maximum value and then decrease with further pressurizing. At about 12 GPa, the superconductivity is completely suppressed (Fig.1d). We repeated the measurements with new samples in five independent experiments and obtained reproducible results.

To know the connection between the superconductivity and the electronic state in FeTe$_{0.55}$Se$_{0.45}$, we performed high-pressure measurements on Hall resistance ($R_{xy}$) by sweeping the magnetic field (*B*), applied perpendicular to the *ab* plane, from 0 T to 2 T on a single crystal sample at various temperatures, as shown in Fig. 2a-e. $R_{xy}(B)$ is negative below 33 K at 0.5 GPa, 28 K at 1.8 GPa, 23 K at 2.4 GPa, respectively. However, at the pressures above 3.4 GPa, the plots of the $R_{xy}(B)$ are positive within the temperature range investigated. These results indicate that an electron-hole carrier balance ($R_{xy}(B)=0$) at the critical temperatures ($T^*$) occurs only below 3.4 GPa. Since the maximum $T_c$ of the pressurized FeTe$_{0.55}$Se$_{0.45}$ is about 22 K, the fixed temperature of 23 K is chosen for the isothermal pressure measurements of the Hall coefficient so as to make a reasonable comparison on the Hall coefficients obtained from different pressures. In this case, the critical pressure where $R_{xy}(B)=0$ is estimated to be ~2.7 GPa (Fig.2f and Fig. S3).

To visualize the correlation between $T_c$ and electronic state in FeTe$_{0.55}$Se$_{0.45}$, we summarize our experimental results in Fig. 3, which demonstrates $R_H$, $T_c$ and structure information of the FeTe$_{0.55}$Se$_{0.45}$ at different pressures. It is seen that $T_c$ is significantly enhanced upon increasing pressure in the pressure range of $0 < P < P_c$ (~2.7 GPa), as shown in the lower panel of Fig.3, while the $R_H$ derived from the Hall resistance $R_{xy}$ becomes less negative (see upper panel of Fig.3 and Fig. S3), reflecting that the contribution of hole carriers to the $T_c$ enhancement is increased.

The connection between the electron state and the lattice structure is one of the key issues for understanding the emergence of the exotic phenomena in correlated electron materials [27, 28]. Interestingly, we noted that the high resolution X-ray diffraction measurements find a temperature-induced structural transition of the tetragonal-orthorhombic (T-O) phase at ~ 40 K in the FeTe$_{0.43}$Se$_{0.57}$ superconductor [24]. Also, a pressure-induced transition from O phase to monoclinic (M) phase was observed in the same sample at ~2.5 GPa below 40 K [24]. Because the composition of the superconductor used for the high pressure XRD measurements is nearly the same as that of our sample, and, in particular, its ambient-pressure transition temperature (~ 40 K) of the T-O phase and the pressure-induced O-M phase transition at low temperature (at ~2.5 GPa) are on the line of our $T^*(P)$ (upper panel of Fig.3), we propose that our samples should share the same structure phase transitions to that of the FeTe$_{0.43}$Se$_{0.57}$ superconductor upon cooling at ambient pressure or at the $P_c$ (~ 2.5 GPa) in the low temperature range. We find that $T^*$ decreases with increasing pressure below $P_c$ (blue region of the upper panel) until undetectable at ~ $P_c$ where the O-M phase transition takes place [24]. This implies that, from ambient pressure to $P_c$, the transport property of the normal state becomes more $p$ type upon increasing pressure. Around the $P_c$, $T_c$ of the orthorhombic superconducting phase reaches to a maximum. On further compression above $P_c$, $T_c$ decreases, while $R_H(P)$ undergoes a sign change from negative to positive, as signified by the change of the color from blue to red (see upper panel of Fig.3 and Fig. S3).

The sign change of $R_H$ in materials is usually associated with a reconstruction of the electronic structure on the Fermi surface (FS) [29-32], so that it can be taken as a fingerprint to manifest the FS reconstruction. Our results demonstrate a close correlation between the FS reconstruction and the T-O or the O-M phase transition. It is interesting to note that the ambient-pressure neutron scattering measurements on superconducting $Fe_{1.08}Te_{0.64}Se_{0.33}$ [33] and $FeTe_{0.5}Se_{0.5}$ [9], whose compositions are similar to that of our sample, show that there are no long-rang magnetic order exist in the samples, but the short-range magnetic correlations with the incommensurate excitation in the superconducting phases. Moreover, angle-resolved photoemission spectroscopy (ARPES) studies found that the normal state of the $FeTe_{0.58}Se_{0.42}$ superconductor presents a strongly correlated metallic feature, which hosts the effective carrier mass up to $16m_e$ [34]. Based on our results and analysis, we propose that the nontrivial superconductivity of this class of materials [11,12] may be associated with the interplay between FS reconstruction and the lattice change, which generates the unusual normal state

In addition, the observed O-M phase transition at the pressure above $P_c$ leads us to propose that the sample may lose its nontrivial superconductivity due to the corresponding change of its crystal structure symmetry needed for protecting the nontrivially topological property [35-38]. Considering no significant change in $R_H(P)$ in the M phase (see upper panel of Fig.3 and Fig.S3), we suggest that the pressure-induced instability, *i.e.* the extent of its lattice distortion, of M phase is responsible for the $T_c$ decrease.

In conclusion, an intimate correlation among the sign change of $R_H$ (a fingerprint for the reconstruction of the Fermi surface), structural phase transition and $T_c$ in the putative topological superconductor FeTe$_{0.55}$Se$_{0.45}$ has been revealed by our high pressure studies for the first time. We find that a noticeable sign change in $R_H$ influences its superconducting transition temperature remarkably. The nontrivially topological superconductivity can be stabilized up to 2.7 GPa ($P_c$), but it may no long exists above $P_c$ due to a crystal structural phase transition. Our results suggest that the nontrivial superconductivity in this material may be associated with its unusual normal state featured by the dramatic interplay between the electronic state and the lattice change. We hope that the correlation among the sign change of $R_H$, structural phase transition and $T_c$ found in this study will shed new light on understanding the entangling state among superconductivity, electronic and lattice structure, and such an entangling state should be responsible for the presence of the nontrivially topological nature of this topological superconductor.

**Acknowledgements**

The work in China was supported by the National Key Research and Development Program of China (Grant No. 2017YFA0302900, 2016YFA0300300 and 2017YFA0303103) and the Strategic Priority Research Program (B) of the Chinese Academy of Sciences (Grant No. XDB25000000). J. G. is grateful for support from the Youth Innovation Promotion Association of the CAS (2019008). Work at Penn State was supported by the US National Science Foundation under grant DMR1707502.

*contributed equally to this work.
†To whom correspondence may be addressed. Email: llsun@iphy.ac.cn.

\#
\#

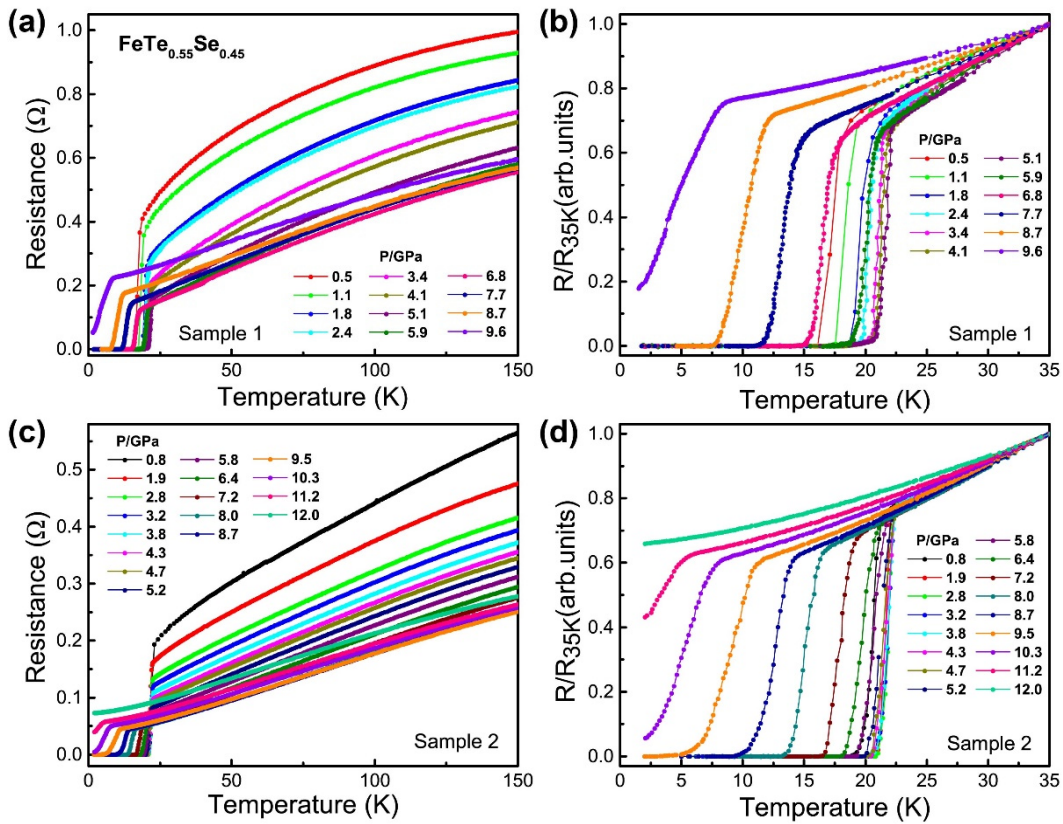

Figure 1. The superconducting behavior of FeTe$_{0.55}$Se$_{0.45}$ at high pressures. (a) Temperature dependence of the resistance in the pressure range of 0.5 GPa–9.6 GPa for the sample1. (b) Enlarged views of the resistance in the lower temperature for the

sample1. (c) Resistance as a function of temperature for pressures ranging from 0.8 GPa to 12 GPa for the sample 2. (d) Resistance versus temperature near the superconducting transition of the sample 2.

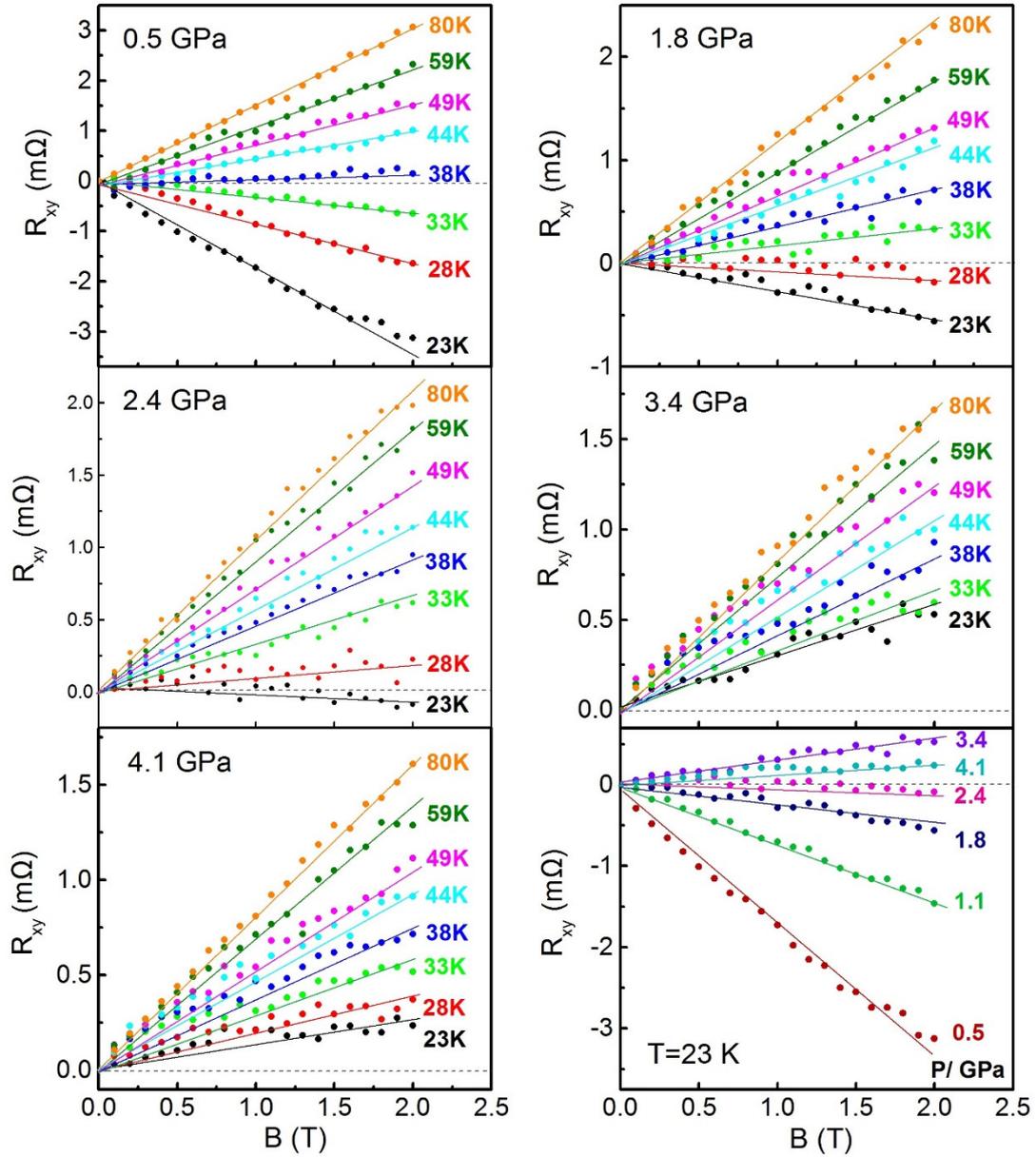

Figure 2 Hall resistance ($R_{xy}$) as a function of magnetic field ($B$) for the FeTe$_{0.55}$Se$_{0.45}$ single crystals. Plots of $R_{xy}$ versus $B$ at different temperatures in the pressure range of (a) 0.5 GPa, (b) 1.8 GPa, (c) 2.4 GPa, (d) 3.4 GPa and (e) 4.1 GPa. (f) $R_{xy}$ versus $B$ at 23 K for pressures ranging from 0.5 GPa to 3.4 GPa. The solid lines are guides to the eye. The dashed line indicates $R_{xy}(B)=0$ where the pressure is estimated to be ~ 2.7 GPa.

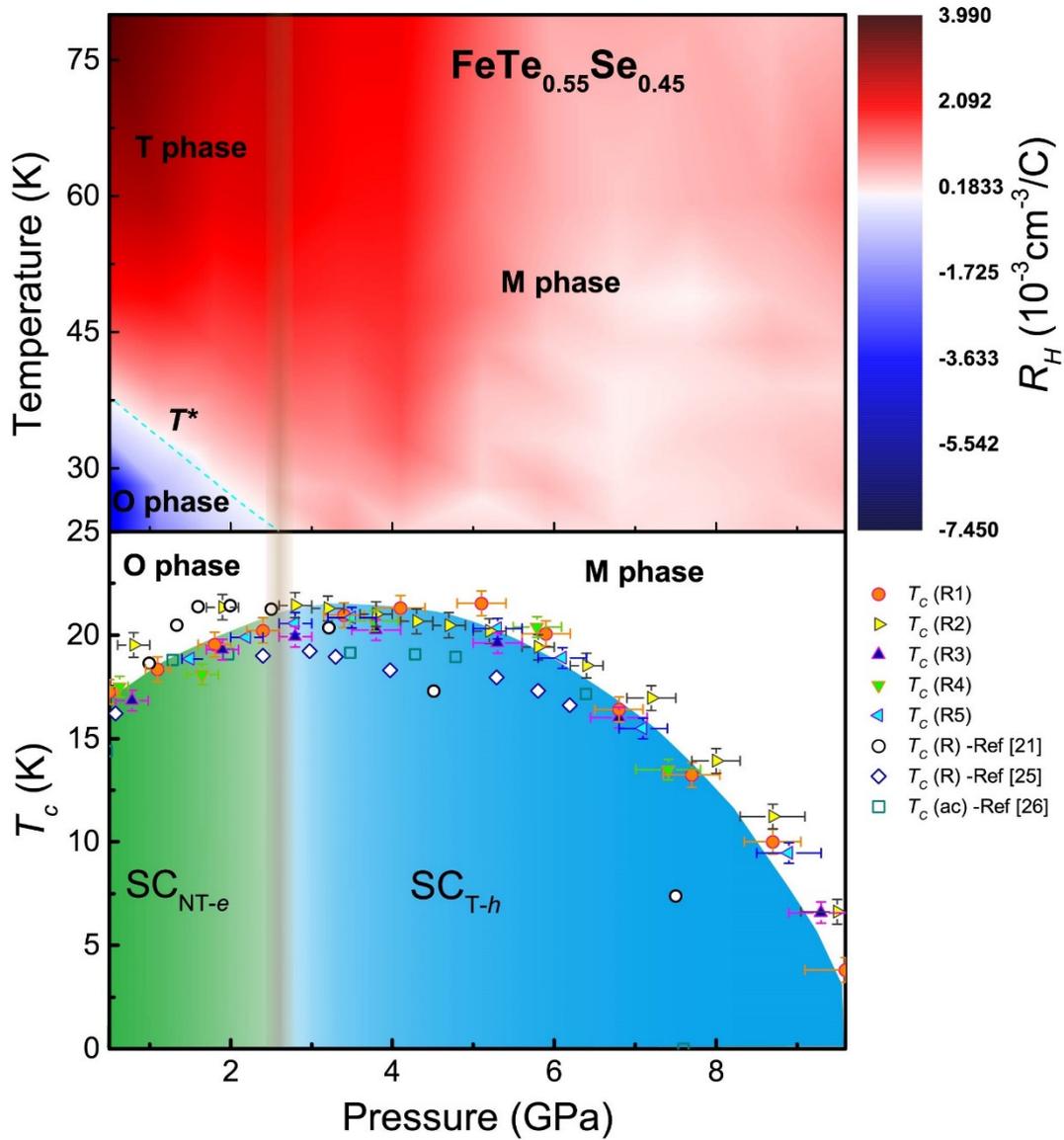

Figure 3 Hall coefficient ($R_H$), structure and superconducting transition temperature ($T_c$) information of the FeTe$_{0.55}$Se$_{0.45}$ superconductor at different pressures. Upper panel presents the mapping information of temperature and pressure dependent $R_H$, shown in color scale. Here $T^*$ represents the temperature of the electron-hole carrier balance. T, O and M stand for the tetragonal, orthorhombic and monoclinic phases, respectively. Lower panel displays $T_c$ as a function of pressure. The values of $T_c$ are determined by the midpoint of the superconducting transition. SC$_{NT}$-$e$ and SC$_T$-$h$ represent the nontrivial superconducting phase with the dominance of electron-carriers and the trivial superconducting phase with the dominance of hole-carriers, respectively. $T_c$(R1), $T_c$(R2), $T_c$(R3), $T_c$(R4) and $T_c$(R5) stand for the $T_c$ obtained by the resistance measurements for the sample 1, sample 2, sample 3, sample 4 and sample 5. $T_c$(ac) and $T_c$(R) represent the $T_c$ obtained by the *ac* susceptibility and resistance measurements.